\documentstyle[aps,prl,manuscript]{revtex} 
\begin{document}
\draft


\title{Electron-electron relaxation in two-dimensional
impure superconductors}

\bigskip
	
\author{Michael Reizer }
\address{5614 Naiche Rd. Columbus OH 43213}

\date{\today}

\maketitle

\begin{abstract}

The electron-electron relaxation in impure two-dimensional 
superconductors is studied.
All channels of the electron-electron
interaction classified in the Nambu representation
are taken into account. 
It is shown that the recombination relaxation rate 
originates from quasipartical processes associated with fluctuations of
the electron density and the phase of the order parameter.
At low temperatures the recombination relaxation rate
has a double exponential temperature dependence.
The scattering relaxation rate at low temperatures has a power
law temperature dependence due to contributions from
gapless collective excitations, the phase modes.
Two-layer superconductor-normal 
metal system is also considered. It is shown that the recombination
relaxation rate in the superconducting layer has a single 
exponential factor at low temperatures  in comparison with a 
one layer superconducting system. This increase in the recombination
relaxation rate originates from the inter-layer Coulomb interaction and 
may be used in constructing of superconducting
radiation detectors.

\end{abstract}
\bigskip

%


\section {Introduction}

The electron-electron energy-relaxation time determines a number of
parameters of nonequilibrium superconductors such as the relaxation
times for the amplitude and the phase of the order parameter \cite{NS}. 
It is also important for nonequilibrium superconducting radiation 
detectors based on the resistive and inductive responses.
The electron-electron relaxation time is responsible for the quasiparticle
multiplication coefficient which in turn determines the responsivity and 
detectivity of the detector and its noise characteristics \cite{SR}.

The energy relaxation time also serves as a pair-breaking parameter in the
superconducting density of states which is measured in the tunneling 
experiment \cite{PL}. Recently the energy relaxation time was measured
in cuprate superconductors by studying an electronic instability
at high vortex velocities in the mixed state \cite{D}.

The electron-electron relaxation time of clean superconductors was
calculated for the three-dimensional case in Ref. \cite{R1} 
and for the two-dimensional case in Ref. \cite{R2}. 
In the last work all channels of the electron-electron
interaction not only the Coulomb interaction was taken into account by 
using the matrix classification of the interaction channels developed
earlier in Refs. \cite{KEO} and \cite{SRW}. 
The importance of considering all channels
of interaction was realized long ago for the problem of gauge invariance
in superconductors \cite{S}. 
It was also emphasized in Ref. \cite{SRW} that interference
between different channels of interaction cancels divergences in the 
interaction correction to the superconducting order parameter.  

It is known that in normal impure and low dimensional 
metals the diffusive motion of 
electrons leads to enhancement of the electron-electron relaxation
\cite{AS}, \cite{AA}.
The purpose of the present paper is to calculate the electron-electron
relaxation time in impure two-dimensional superconductors using the 
formalism of Refs. \cite{KEO} and \cite {SRW}. Earlier attempt to study 
electron-electron relaxation in impure two-dimensional
superconductors \cite{DB}
took into account only the Coulomb electron-electron interaction and 
therefore ignored the other relevant interaction 
channels. As we will show in the present paper including all channels of the
interaction is very important for the electron-electron relaxation,
and leads to results qualitatively different from that of Ref. \cite{DB}. 
Another difference is that we found important contributions to
the scattering relaxation time from gapless collective excitations
which where mised in Ref. \cite{DB}.
Note also that we derived the quasiparticle energy relaxation time from 
the quantum kinetic equation, not as an imaginary part of the self-energy
as in Ref. \cite{DB}.

Then we study the superconductor-normal metal two-layer system.
Such a system was already studied in Ref. \cite{R2} for a clean system.
We consider a disordered case in the present work and show that
recombination relaxation rate is strongly enhanced due to inter-layer
electron-electron interaction.   

\bigskip

\section {Electron-electron interaction in matrix formalism}

\bigskip

We use the Keldysh diagram technique for nonequilibrium processes in 
which the electron Green's functions, along with the electron-electron
interaction potential, the electron 
self energy and the polarization operator are represented by supermatrices
\begin{eqnarray}
\label{1}
({\hat G})=\pmatrix{0&{\hat G}^A\cr {\hat G}^R&{\hat G}^C},
\ \ \ ({\hat V})= \pmatrix{0&{\hat V}^A\cr {\hat V}^R&{\hat V}^C},\cr
({\hat \Sigma})=\pmatrix{{\hat \Sigma}^C&{\hat \Sigma}^R\cr 
{\hat\Sigma}^A&0},\ \ \  
({\hat \Pi})=\pmatrix{{\hat \Pi}^C&{\hat \Pi}^R\cr {\hat \Pi}^A&0}.  
\end{eqnarray}   
The matrix electron Green's function in an impure superconductor in the Nambu
representation has the form
\begin{eqnarray}
\label{2}
\hat G^R(P)=[\hat G^A(P)]^*=
{-\xi_p\hat\tau_3-\epsilon^R\hat\tau_0+\Delta^R\hat\tau_1
\over \xi_p^2-(E^R)^2},\cr             
P=({\bf p},\epsilon),\ \ \  \xi_p={p^2-p_F^2\over 2m},       
\end{eqnarray}
where $\hat \tau_i$ are the Pauli matrices, 
$m$ is the electron mass, and 
\begin{eqnarray}
\label{3}
\epsilon^R=\epsilon\biggl(1+{i\over 2\tau\xi_\epsilon}\biggr),\ \ \ 
\Delta^R=\Delta\biggl(1+{i\over 2\tau\xi_\epsilon}\biggr),\cr 
(E^R)=(\epsilon^R)^2+(\Delta^R)^2,    
\end{eqnarray}                               
where $\Delta$ is the energy gap, $\tau$ is the electron-impurity 
relaxation time, and
\begin{eqnarray}
\label{4}
\xi_\epsilon=(\epsilon^2-\Delta^2)^{1/2}{\rm sgn}(\epsilon), 
\ \ \  |\epsilon|>\Delta.                                           
\end{eqnarray}
In a spatially uniform system the kinetic components $\hat G^C$ and 
$\hat\Sigma^C$
are satisfied the equations
\begin{eqnarray}
\label{5}
\hat G^C(P)=S(\epsilon)[\hat G^A(P)-\hat G^R(P)],\cr 
\hat\Sigma(P)=S(\epsilon)[\hat \Sigma^A(P)-\hat \Sigma^R(P)],       
\end{eqnarray}
where $S(\epsilon)=-\tanh(\epsilon/2T)=2n(\epsilon)-1$, and $n(\epsilon)$
is the Fermi distribution function.
Similar relations hold for matrix interaction potentials and
polarization operators:
\begin{eqnarray}
\label{6}
\hat V^C(Q)=(2N(\omega)+1)[\hat V^R(Q)-\hat V^A(Q)],\cr
\hat\Pi^C(Q)=(2N(\omega)+1)[\hat\Pi^R(Q)-\hat\Pi^A(Q)],\ \ 
\end{eqnarray}                                                 
where $Q=({\bf q},\omega)$ 
and $N(\omega)$ is the Bose distribution function.

For averaging over impurity position it is convenient to introduce the 
following expressions
\begin{eqnarray}
\label{7}
\eta^{AA}_i={1\over \pi\nu\tau}<\hat\tau_3\hat G^A(P)
\hat\tau_i\hat G^A(P+Q)\hat\tau_3>,\cr
\eta^{AR}_i={1\over \pi\nu\tau}<\hat\tau_3\hat G^A(P)
\hat\tau_i\hat G^R(P+Q)\hat\tau_3>, \cr
<...>=\int {d^2p\over (2\pi)^2},
\end{eqnarray}                                                                 
where $\nu$ is the two-spin electron density of states. 
Calculations give e.g.
\begin{eqnarray}
\label{8}
\eta_0^{AR}={\zeta\over 2}[(1+A_+)\hat\tau_0+B_+\hat\tau_1], \cr
\eta_3^{AA}={\zeta_+\over 2}[(1-A)\hat\tau_3+Bi\hat\tau_2],  
\end{eqnarray}
where
\begin{eqnarray}
\label{9}
A={\epsilon(\epsilon+\omega)-\Delta^2\over 
\xi_\epsilon\xi_{\epsilon+\omega}},\ \ \ 
A_+={\epsilon(\epsilon+\omega)+\Delta^2\over 
\xi_\epsilon\xi_{\epsilon+\omega}},\cr                                      
B={(\epsilon+\omega)\Delta-\epsilon\Delta\over 
\xi_\epsilon\xi_{\epsilon+\omega}}={\omega\Delta\over
\xi_\epsilon\xi_{\epsilon+\omega}},\cr 
B_+={(\epsilon+\omega)\Delta+\epsilon\Delta\over 
\xi_\epsilon\xi_{\epsilon+\omega}}={(2\epsilon+\omega)\Delta\over
\xi_\epsilon\xi_{\epsilon+\omega}},
\end{eqnarray}                                
\begin{eqnarray}
\label{10}
\zeta=1+i(\xi_{\epsilon+\omega}-\xi_{\epsilon})\tau-Dq^2\tau,\cr
\zeta_+=1-i(\xi_{\epsilon+\omega}+\xi_{\epsilon})\tau-Dq^2\tau. 
\end{eqnarray}
Values of $\eta_i$ for different matrices $\hat\tau_i$ are presented in Tab. 1.
Note also that the following identities hold: $A^2-B^2=1$ and $A_+^2-B_+^2=-1$.

Treating the electron-electron interaction in superconductors we use the matrix
formalism developed in Refs. 7 and 8.
The bare vertices for the electron-electron interaction are
classified in terms of the Pauli matrices. Physical meaning of the 
corresponding operators $\hat O_i=\Psi^\dagger\hat\tau_i\Psi$ is the following.
Matrix $\hat\tau_1$ corresponds to the order parameter amplitude $\Delta$, 
matrix $\hat\tau_2$ corresponds to the order parameter phase $\phi$, 
matrix $\hat\tau_3$ corresponds to the electron density, and the vector
matrix ${\bf k}\hat\tau_0$ corresponds to the electric current, the later
will not be considered in the present paper(see Ref. 5). 
Note also that each impurity vertex carries the matrix $\hat\tau_3$.

Therefore each interaction vertex operates in both Keldysh and Nambu
spaces and has the form $\hat\gamma^k_{mn}(\hat\tau_i)=
\hat\gamma(\hat\tau_i)K^k_{mn}$, where $\hat\tau_i$
indicates the component in the Nambu space and tensor $K^k_{mn}$
stands for the Keldysh space, $k$ is a boson index and $m$ and $n$ are
the electron indices. In the representation corresponding to Eq. 1 the nonzero
components of tensor $K^k_{mn}$ are
\begin{eqnarray}
\label{11}
K^1_{22}=K^1_{11}=K^2_{12}=K^2_{21}={1\over {\sqrt 2}}.
\end{eqnarray}           
We will omit coefficient $1/{\sqrt 2}$ 
in intermediate equations for the impurity renormalized vertices and
restore it in final equations for the polarization operators and the
electron self energies. 

Impurity averaging leads to the  ladder equation for the scalar vertex 
$\hat \Gamma(\hat\tau_i)$ 
shown in Fig. 1. Such an equation should be written for each bare 
matrix $\hat\tau_i$ in the Nambu space.
The solution of these equations for the vertex $\hat \Gamma(\hat\tau_i)$
in the Keldysh-Nambu space is obtained following Ref. 13.
We start with the equation for the vertex 
$\hat \Gamma^1_{22}(\hat\tau_3)$,
\begin{eqnarray}
\label{12}
<\hat\tau_3\hat G^A(P)\hat \Gamma^1_{22}(\hat\tau_3)\hat G^A(P+Q)\hat\tau_3>
-\hat \Gamma^1_{22}(\hat\tau_3)=-\hat\tau_3.     
\end{eqnarray}
The solution of this equation is
\begin{eqnarray}
\label{13}
\hat \Gamma^1_{22}(\hat\tau_3)=-{\zeta\over 2}{B\over 1-\zeta}i\hat\tau_2+
\biggl(1+{\zeta\over 2}{1+A\over 1-\zeta}\biggr)\hat\tau_3.        
\end{eqnarray}
Note that renormalized vertex $\hat \Gamma^1_{22}(\hat\tau_3)$ has components
proportional not only to to matrix $\hat\tau_3$ but also to matrix 
$\hat\tau_2$. 
The other vertices
$\hat \Gamma^1_{22}(\hat\tau_i)$ and $\hat \Gamma^2_{21}(\hat\tau_i)$ are
presented in Tab. 2. The vertices with the other Keldysh indices are
obtained from the equations
\begin{eqnarray}
\label{14}
\hat \Gamma^2_{12}(\hat\tau_i)=(\hat \Gamma^2_{21}(\hat\tau_i))^*,
\ \ \hat \Gamma^1_{12}(\hat\tau_i)=S(\epsilon)[\hat \Gamma^1_{22}(\hat\tau_i)-
\hat \Gamma^2_{12}(\hat\tau_i)],\cr
\hat \Gamma^1_{21}(\hat\tau_i)=-S(\epsilon+\omega)
[\hat \Gamma^1_{22}(\hat\tau_i)-
\hat \Gamma^2_{21}(\hat\tau_i)],\cr
\hat \Gamma^2_{11}(\hat\tau_i)=S(\epsilon)\hat \Gamma^2_{21}(\hat\tau_i)
-S(\epsilon+\omega)\hat \Gamma^2_{12}(\hat\tau_i)\cr
-[S(\epsilon)-S(\epsilon+\omega)](\hat\Gamma^1_{22}(\hat\tau_i))^*.     
\end{eqnarray}    
                                                                 
Note also that index structure of the renormalized vertices in the Keldysh
space is different from the index structure of the bare vertex described by
Eq. (11). It is important that the complex conjugate in Eq. (14) operates 
only in the Keldysh space, thus any $i$ in Tab. 2 originating from $\hat\tau_i$
matrix algebra are not affected by the complex conjugate, e. g. (compare with
Eq. (13))
\begin{eqnarray}
\label{15}
(\hat \Gamma^1_{22}(\hat\tau_3))^*
=-{\zeta^*\over 2}{B\over 1-\zeta^*}i\hat\tau_2+
\biggl(1+{\zeta^*\over 2}{1+A\over 1-\zeta^*}\biggr)\hat\tau_3.
\end{eqnarray}
The effective screened electron-electron interaction in a superconductor 
according to Refs. 6-8 is shown in Fig. 1. The solution of this equation
in $3\times 3$ matrix form is
\begin{eqnarray}
\label{16}
\hat V=\pmatrix{-(2/\lambda+\Pi_{11})^{-1}&0&0
\cr 0&[(V_0)^{-1}-\Pi_{33}]/\cal D & 
\Pi_{23}/\cal D\cr 0& \Pi_{32}/\cal D & -(2/\lambda+\Pi_{22})/\cal D},      
\end{eqnarray}                                                                 
where
\begin{eqnarray}
\label{17}
{\cal D}=-(2/\lambda+\Pi_{22})[(V_0)^{-1}-\Pi_{33}]-
\Pi_{23}\Pi_{32},                                               
\end{eqnarray}
and $\lambda$ is the BCS coupling constant ($\lambda>0$), 
$V_0=2\pi e^2/q$ is the nonscreened two-dimensional Coulomb potential. 
                                 
The polarization operators renormalized by impurities are expressed 
through the vertex $\hat \Gamma$ by the equation
\begin{eqnarray}
\label{18}
\Pi^A_{ij}(Q)=-{i\over 2}\pi\nu\tau\int{d \epsilon\over 2\pi}
{\rm Sp}\biggl(\hat\tau_3\hat\tau_j\hat\tau_3\Gamma^2_{11}(\hat\tau_i)
\biggr).                                                        
\end{eqnarray}
Using Eq. (14) for $\Gamma^2_{11}(\hat\tau_i)$ and Tab. 2  we 
find the polarization operators,
\begin{eqnarray}
\label{19}
\Pi^A_{11}(Q)={i\nu\tau\over 4}\int d\epsilon
\biggl((1-A_+){S(\epsilon+\omega)-S(\epsilon)\over 1-\zeta^*}-
(1+A_+)\biggl[{S(\epsilon+\omega)\over 1-\zeta_+^*}
-{S(\epsilon)\over 1-\zeta_+}\biggr]\biggr),                       
\end{eqnarray}
\begin{eqnarray}
\label{20}
\Pi^A_{22}(Q)={i\nu\tau\over 4}\int d\epsilon
\biggl((1-A){S(\epsilon+\omega)-S(\epsilon)\over 1-\zeta^*}-
(1+A)\biggl[{S(\epsilon+\omega)\over 1-\zeta_+^*}
-{S(\epsilon)\over 1-\zeta_+}\biggr]\biggr),                        
\end{eqnarray}
\begin{eqnarray}
\label{21}
\Pi^A_{33}(Q)=-\nu-{i\nu\tau\over 4}\int d\epsilon
\biggl((1+A){S(\epsilon+\omega)-S(\epsilon)\over 1-\zeta^*}-
(1-A)\biggl[{S(\epsilon+\omega)\over 1-\zeta_+^*}
-{S(\epsilon)\over 1-\zeta_+}\biggr]\biggr),                        
\end{eqnarray}
\begin{eqnarray}
\label{22}
\Pi^A_{32}(Q)=-\Pi^A_{23}(Q)=-{\nu\tau\over 4}\int d\epsilon B
\biggl({S(\epsilon)-S(\epsilon+\omega)\over 1-\zeta^*}
-{S(\epsilon+\omega)\over 1-\zeta_+^*}
+{S(\epsilon)\over 1-\zeta_+}\biggr).                              
\end{eqnarray}
To calculate the electron relaxation we need 
the imaginary part of the potentials (the polarization operators)
in the quasiparticle
representation. Making a transformation from the electronic
representation to the quasiparticle representation we use Eq. (4).
As a result we separate out the processes of scattering and recombination of 
quasiparticles in Eqs. (18)-(21).
For the imaginary part of the polarization operators we have
\begin{eqnarray}
\label{23}
{\rm Im}\Pi^A_{ii}(Q)_{scatt}={\nu\over 2}
\int_\Delta^\infty d\epsilon
[S(\epsilon+\omega)-S(\epsilon)]C_{ii}(q,\epsilon,\omega), 
\end{eqnarray}
\begin{eqnarray}
\label{24}
{\rm Im}\Pi^A_{ii}(Q)_{recom}={\nu\over 4}\Theta(\omega-2\Delta)
\int_\Delta^{\omega-\Delta}d\epsilon
[S(\epsilon-\omega)-S(\epsilon)]C_{ii}(q,\epsilon,-\omega),     
\end{eqnarray}
where $C_{ii}$ are
\begin{eqnarray}
\label{25} 
C_{11}(q,\epsilon,\omega)=
\biggl(1-{\epsilon(\epsilon+\omega)+\Delta^2\over 
|\xi_\epsilon||\xi_{\epsilon+\omega}|}\biggr)
{Dq^2\over (|\xi_{\epsilon+\omega}|-|\xi_\epsilon|)^2+(Dq^2)^2}\cr
-\biggl(1+{\epsilon(\epsilon+\omega)+\Delta^2\over
|\xi_\epsilon||\xi_{\epsilon+\omega}|}\biggr)
{Dq^2\over (|\xi_{\epsilon+\omega}|+|\xi_\epsilon|)^2+(Dq^2)^2},
\end{eqnarray}
\begin{eqnarray}
\label{26}
C_{22}(q,\epsilon,\omega)=
\biggl(1-{\epsilon(\epsilon+\omega)-\Delta^2\over
|\xi_\epsilon||\xi_{\epsilon+\omega}|}\biggr)
{Dq^2\over (|\xi_{\epsilon+\omega}|-|\xi_\epsilon|)^2+(Dq^2)^2}\cr
-\biggl(1+{\epsilon(\epsilon+\omega)-\Delta^2
\over|\xi_\epsilon||\xi_{\epsilon+\omega}|}\biggr)
{Dq^2\over (|\xi_{\epsilon+\omega}|+|\xi_\epsilon|)^2+(Dq^2)^2} , 
\end{eqnarray}
\begin{eqnarray}
\label{27}
C_{33}(q,\epsilon,\omega)=
-\biggl(1+{\epsilon(\epsilon+\omega)-\Delta^2\over
|\xi_\epsilon||\xi_{\epsilon+\omega}|}\biggr)
{Dq^2\over (|\xi_{\epsilon+\omega}|-|\xi_\epsilon|)^2+(Dq^2)^2}\cr
+\biggl(1-{\epsilon(\epsilon+\omega)-\Delta^2
\over|\xi_\epsilon||\xi_{\epsilon+\omega}|}\biggr)
{Dq^2\over (|\xi_{\epsilon+\omega}|+|\xi_\epsilon|)^2+(Dq^2)^2}. 
\end{eqnarray}

For the off-diagonal polarization operator we have
\begin{eqnarray}
\label{28}
\Pi^A_{32}(Q)_{scatt}=-{\nu\over 2}\Delta\omega
\int_\Delta^\infty {d\epsilon\over|\xi_\epsilon||\xi_{\epsilon+\omega}|}
\biggl[{(S(\epsilon)-S(\epsilon+\omega))\over
i|\xi_{\epsilon+\omega}|-i|\xi_\epsilon|+Dq^2}\cr
+{S(\epsilon)\over i|\xi_{\epsilon+\omega}|+i|\xi_\epsilon|+Dq^2}
-{S(\epsilon+\omega)\over -i|\xi_{\epsilon+\omega}|
-i|\xi_\epsilon|+Dq^2}\biggr],                                    
\end{eqnarray}
\begin{eqnarray}
\label{29}
\Pi^A_{32}(Q)_{rec}=-\Theta(\omega-2\Delta){\nu\over 2}\Delta\omega
\int_\Delta^{\omega-\Delta} {d\epsilon\over
|\xi_\epsilon||\xi_{\epsilon-\omega}|}\biggl[
{S(\epsilon)-S(\epsilon-\omega)\over 
i|\xi_{\epsilon-\omega}|-i|\xi_\epsilon|+Dq^2}\cr
+{S(\epsilon)\over i|\xi_{\epsilon-\omega}|+i|\xi_\epsilon|+Dq^2}
-{S(\epsilon-\omega)\over -i|\xi_{\epsilon-\omega}|
-i|\xi_\epsilon|+Dq^2}\biggr].                                   
\end{eqnarray}

As we will see in the next chapter, 
calculating the electron relaxation time 
we need the imaginary part of the propagators ${\rm Im} (V^A_{ii}(Q))$.
The imaginary part of the propagators may originate from the poles
of the propagators which correspond to collective excitations or
from the imaginary part of the polarization operators
${\rm Im} (\Pi^A_{ii}(Q))$, which correspond to real
processes of scattering and recombination of quasiparticles.
We will restrict our calculations to low temperatures $T<<\Delta$,
where large frequencies $\omega-2\Delta<<T$, 
are important for both recombination and scattering
processes and small frequencies 
$\omega<<T<<\Delta$ are important only for the scattering processes.
The imaginary part of diagonal polarization operators 
for in these regions are presented in Tab. 3. 
Real parts of the polarization operators are analyzed in Appendix A.

Now we study in detail each of the matrix elements of $\hat V$.
Following Eq. (A17) the imaginary part of the potential $V_{11}$ 
for scattering processes and for small arguments $\omega<<\Delta$ and
$Dq^2<<\Delta$ is
\begin{eqnarray}
\label{30}
{\rm Im}V_{11}^A(Q)=-{\rm Im}{1\over 2/\lambda+\Pi_{11}^A(Q)}\cr
\approx{\rm Im}{1\over \nu/2-i{\rm Im}\Pi_{11}^A(Q)}
\approx\biggl({1\over \nu}\biggr)^2{\rm Im}\Pi_{11}^A(Q).   
\end{eqnarray}
This approximation is justified because there is no singularity
in the order parameter amplitude propagator $V_{11}$ at small frequency
and momentum for any finite $\Delta$, which means that fluctuations of
the amplitude of the order-parameter are massive. 
For the recombination processes, large frequencies $\omega>2\Delta$ are
important, and according to Tab. 3 ${\rm Im}\Pi_{11}$ may be neglected.

As was shown in Appendix A, the propagators $V_{33}$ and 
$V_{22}$ for $\omega<<\Delta$ and $Dq^2<<\Delta$ have the form
\begin{eqnarray}
\label{31}
V^A_{33}(Q)={\kappa\over \nu q}{\pi\Delta Dq^2-\omega^2\over  
\pi\Delta Dq\kappa-(\omega-i0)^2},                           
\end{eqnarray}
\begin{eqnarray}
\label{32}
V^A_{22}(Q)={\kappa\over\nu  q}{4\Delta^2\over  
\pi\Delta Dq\kappa-(\omega-i0)^2}{\pi\Delta Dq^2-\omega^2\over 
\pi\Delta Dq^2+\omega^2}.                          
\end{eqnarray}
Thus the imaginary part of the propagators $V_{33}$ and 
$V_{22}$ for small arguments comes from the pole corresponding to 
the phase mode, not from the imaginary part of the 
polarization operators. 

For large frequencies $\omega\sim 2\Delta$ we use the
following approximation, 
\begin{eqnarray}
\label{33}
{\rm Im}V^A_{33}(Q)
=\biggl(V^{-1}_0(q)-\Pi_{33}^A(Q)-
{(\Pi_{23}^3(Q))^2\over 2/\lambda+\Pi_{22}^3(Q)}\biggr)^{-1}
\approx V_S(q)^2{\rm Im}\Pi^A_{33}(Q),                        
\end{eqnarray}
\begin{eqnarray}
\label{34}
{\rm Im}V_{22}^A(Q)
={\rm Im}\biggl( -{2\over \lambda}-\Pi_{22}^A(Q)+
{(\Pi_{23}^A(Q))^2\over V_0(q)^{-1}
-\Pi_{33}^A(Q)}\biggr)^{-1}\approx
{\lambda^2\over 4}{\rm Im}\Pi_{22}^A(Q).                      
\end{eqnarray}
where $V_S(q)=\kappa/\nu(q+\kappa)$ is the statically screened Coulomb 
potential in the normal state,
which was presented above for the two-dimensional case, $\kappa=2\pi\nu e^2$
is the screening momentum.
Such an approximation is justified due to absence of
collective excitation in this frequency region.

\bigskip

\section{Electron-electron relaxation}

\bigskip
             
The kinetic equation for nonequilibrium distribution function in a
spatially uniform system is
\begin{eqnarray}
\label{35}
{d n(\epsilon)\over d t}=-{i\over \pi\nu}{\xi_\epsilon\over \epsilon}
{1\over 2}{\rm Tr}\int{d^2p\over (2\pi)^2}
{\rm Im}[\hat G^A(P)][\hat\Sigma^C(P)-S(\epsilon)(\hat\Sigma^A(P)
-\hat\Sigma^R(P))].     
\end{eqnarray}
The electron energy relaxation time $\tau_{e-e}$ is 
determined from the equation
\begin{eqnarray}
\label{36}
{1\over \tau_{e-e}(T,\epsilon)}=-{\partial \over \partial n(\epsilon)}
{dn(\epsilon)\over d t}.                                           
\end{eqnarray}
The electron self-energy is shown in Fig. 3.
Using the results of Section 2, we have
\begin{eqnarray}
\label{37}
{1\over \tau_{e-e}(T,\epsilon)}=
{2\over \pi }{\xi_\epsilon\over \epsilon}
\int {dQ\over (2\pi)^3}[N(\omega)+n(\omega+\epsilon)]\cr
\times{\delta\over \delta S(\epsilon)}
\lbrace{\rm Im}V^A_{ii}(Q){\rm Re}[{\rm Pr}_{\hat\tau_i}(\hat\Gamma^2_{11}
(\hat\tau_i))]-2{\rm Im}V^A_{23}{\rm Im}[{\rm Pr}_{\hat\tau_3}
(\hat\Gamma^2_{11}(\hat\tau_2))]\rbrace,                          
\end{eqnarray}
where ${\rm Pr}_{\hat\tau_i}$ means the component proportional to matrix 
$\hat\tau_i$ (projection on $\hat\tau_i$). In Eq. (37) summation on 
repeated indices is implied. 
Using Tab. 2 and relation $\delta \Gamma^2_{11}/\delta S(\epsilon)=
\Gamma^2_{21}-(\Gamma^1_{22})^*$ we present Eq. (37) in the form
\begin{eqnarray}
\label{38}
{1\over \tau_{e-e}(T,\epsilon)}={1\over \pi^2}\int d\omega
\int {d^2q\over (2\pi)^2}[N(\omega)+n(\omega+\epsilon)]\cr
\times\biggl[{\rm Im}V_{11}^A(Q)
\biggl(\biggl({\xi_\epsilon\over \epsilon}
+{\epsilon(\epsilon+\omega)+\Delta^2\over 
\epsilon\xi_{\epsilon+\omega}}\biggr)
{Dq^2\over (\xi_{\epsilon+\omega}+\xi_\epsilon)^2+(Dq^2)^2}\cr
-\biggl({\xi_\epsilon\over \epsilon}
-{\epsilon(\epsilon+\omega)+\Delta^2\over 
\epsilon\xi_{\epsilon+\omega}}\biggr)
{Dq^2\over (\xi_{\epsilon+\omega}-\xi_\epsilon)^2+(Dq^2)^2}\biggr)\cr
+{\rm Im}V_{22}^A(Q)
\biggl(\biggl({\xi_\epsilon\over \epsilon}
+{\epsilon(\epsilon+\omega)-\Delta^2\over 
\epsilon\xi_{\epsilon+\omega}}\biggr)
{Dq^2\over (\xi_{\epsilon+\omega}+\xi_\epsilon)^2+(Dq^2)^2}\cr
-\biggl({\xi_\epsilon\over \epsilon}
-{\epsilon(\epsilon+\omega)-\Delta^2\over 
\epsilon\xi_{\epsilon+\omega}}\biggr)
{Dq^2\over (\xi_{\epsilon+\omega}-\xi_\epsilon)^2+(Dq^2)^2}\biggr)\cr
+{\rm Im}V_{33}^A(Q)
\biggl(\biggl({\xi_\epsilon\over \epsilon}
-{\epsilon(\epsilon+\omega)-\Delta^2\over 
\epsilon\xi_{\epsilon+\omega}}\biggr)
{Dq^2\over (\xi_{\epsilon+\omega}+\xi_\epsilon)^2+(Dq^2)^2}\cr
-\biggl({\xi_\epsilon\over \epsilon}
+{\epsilon(\epsilon+\omega)-\Delta^2\over 
\epsilon\xi_{\epsilon+\omega}}\biggr)
{Dq^2\over (\xi_{\epsilon+\omega}-\xi_\epsilon)^2+(Dq^2)^2}\biggr)\cr
+2i{\rm Im}V^A_{23}{\omega\Delta\over\epsilon\xi_{\epsilon+\omega}}
\biggl({Dq^2\over (\xi_{\epsilon+\omega}+\xi_\epsilon)^2+(Dq^2)^2}
+{Dq^2\over (\xi_{\epsilon+\omega}-\xi_\epsilon)^2+(Dq^2)^2}\biggr)\biggr].
\end{eqnarray}

In order to separate out the processes of scattering and recombination of 
quasiparticles we need to make a transformation from the electronic
representation to the quasiparticle representation in Eq. (38).
Note that the presence of imaginary factor $i$ in the last term in Eq. (38) 
means that for the contribution of the nondiagonal channels of
interaction requires the states under the gap to be taken into account
according to the equation: $\zeta_{\epsilon+\omega}=i[\Delta^2-
(\omega+\epsilon)^2]^{1/2},\ \ |\epsilon+\omega|<\Delta$. Such states 
should also be included in equations for the polarization operator
$\Pi_{23}$, in Eqs. (28) and (29) only the states above the gap were
included. The analysis similar to that presented in Ref. 6 for the clean 
case shows that contribution from the nondiagonal channels of interaction
may be neglected.

For electrons on the Fermi surface, $\epsilon=\Delta$,
\begin{eqnarray}
\label{39}
{1\over \tau_{e-e}(T,\epsilon=\Delta)}_{scatt}=
{2\over \pi^3}\int_{2\Delta}^\infty 
d\omega\int_0^\infty dqq[N(\omega)+n(\omega-\Delta)]
{Dq^2\over \omega^2-\Delta\omega +(Dq^2)^2}\cr
\times\biggl[[{\rm Im}V_{22}^A(Q)_{recom}
+{\rm Im}V_{33}^A(Q)_{recom}]
\biggl({\Delta\over 2\omega}\biggl)^{1/2}\biggr]\cr
+{2\over \pi^3}\int_0^\infty 
d\omega\int_0^\infty dqq[N(\omega)+n(\omega+\Delta)]
{Dq^2\over 2\Delta\omega+\omega^2 +(Dq^2)^2}\cr
\times\biggl[{\rm Im}V_{11}^A(Q)_{scatt}
\biggl({2\Delta+\omega\over \omega}\biggr)^{1/2}+[{\rm Im}V_{22}^A(Q)_{scatt}+
{\rm Im}V_{33}^A(Q)_{scatt}]
\biggl({\Delta\over 2\omega}\biggl)^{1/2}\biggr],
\end{eqnarray}
\begin{eqnarray}
\label{40}
{1\over \tau_{e-e}(T,\epsilon=\Delta)}_{recom}=
{2\over \pi^3}\int_{2\Delta}^\infty 
d\omega\int_0^\infty dqq[N(\omega)+n(\omega-\Delta)]
{Dq^2\over \omega^2-\Delta\omega +(Dq^2)^2}\cr
\times\biggl[{\rm Im}V_{11}^A(Q)_{scatt}
\biggl({\omega-2\Delta\over \omega}\biggr)^{1/2}+[{\rm Im}V_{22}^A(Q)_{scatt}
+{\rm Im}V_{33}^A(Q)_{scatt}]
\biggl({\Delta\over 2\omega}\biggl)^{1/2}\biggr]\cr
+{2\over \pi^3}\int_{2\Delta}^\infty 
d\omega\int_0^\infty dqq[N(\omega)+n(\omega+\Delta)]
{Dq^2\over 2\Delta\omega+\omega^2 +(Dq^2)^2}\cr
\times\biggl[[{\rm Im}V_{22}^A(Q)_{recom}+
{\rm Im}V_{33}^A(Q)_{recom}]
\biggl({\Delta\over 2\omega}\biggl)^{1/2}\biggr].                 
\end{eqnarray}
Eqs. (39) and  (40) describe processes of scattering ``two into two''
and ``three into one'' quasiparticles  correspondingly.   

Further calculations will be performed for low temperatures $T<<\Delta$.
It may be shown that the most important contribution to the recombination time 
originates from terms $V_{22}(Q)_{recom}$ and $V_{33}(Q)_{recom}$ 
in Eq. (40),
\begin{eqnarray}
\label{41}
{1\over \tau_{e-e}(T,\epsilon=\Delta)}_{recom}=
{T\over 4\pi D\nu}\biggl(1+\biggl({\lambda\nu\over 2}\biggr)^2\biggr)
\exp\biggl(-{2\Delta\over T}\biggr).                              
\end{eqnarray}
Calculating the scattering relaxation time from terms
$V_{22}(Q)_{recom}$ and $V_{33}(Q)_{recom}$ in Eq. (38) we
use the approximation of Eqs. (33) and (34) and we use
Tab. 3 for the imaginary parts of the 
polarization operators. As a result we get
\begin{eqnarray}
\label{42}
{1\over \tau_{e-e}(T,\epsilon=\Delta)}_{scatt}=
{T\over 2\pi^2D\nu}  \biggl(1+
\biggl({\lambda\nu\over 2}\biggr)^2\biggl)
\exp\biggl(-{\Delta\over T}\biggr).                             
\end{eqnarray}

As for the contribution to the scattering relaxation time from terms 
${\rm Im}V_{22}(Q)_{scatt}$ and ${\rm Im}V_{33}(Q)_{scatt}$ in Eq. (38)
we note that for small energy transfers, $\omega<<T<<\Delta$
the imaginary part of the propagators $V_{33}$ and $V_{22}$ originates from
the poles corresponding to the phase mode as seen in Eqs. (31) and (32)). 
Integrating these poles over the momentum $q$ we get
\begin{eqnarray}
\label{43}
{1\over \tau_{e-e}(T,\epsilon=\Delta)}_{scatt}=
{21\over 2\pi^6}\biggl({\pi\over 2}\biggr)^{1/2}
\biggl({T\over \Delta}\biggr)^{3/2}
{T^2\over D^2\kappa^2\nu}.
\end{eqnarray}
Note that the main contribution comes from the propagator $V_{22}$
corresponding to the fluctuation of the phase of the order parameter.

The term $V_{11}(Q)_{scatt}$ does not have poles corresponding to 
the collective mode, thus using Eq. (30) we find that
the scattering relaxation time has  
a power law divergence,
\begin{eqnarray}
\label{44}
{1\over \tau_{e-e}(T,\epsilon=\Delta)}_{scatt}=
{\Delta\over 2\pi^2D\nu} 
\biggl({\pi T\over \omega_0}\biggr)^{1/2}
\exp\biggl(-{\Delta\over T}\biggr).                             
\end{eqnarray}
This divergence is similar to the logarithmic divergence of the 
phase or energy relaxation times in the normal impure two-dimensional 
case$^{14}$.
According to Ref. 14 the cutoff frequency $\omega_0$ is defined by the 
relaxation time
$\tau_{e-e}$, which physically means that the kinetic equation cannot
be applied for energy transfers less than $1/\tau_{e-e}$, thus 
the self-consistent solution of Eq. (44) is
\begin{eqnarray}
\label{45}
{1\over \tau_{e-e}(T,\epsilon=\Delta)}_{scatt}=
{(\pi\Delta^2T)^{1/3}\over (2\pi^2D\nu)^{3/2}}
\exp\biggl(-{2\Delta\over 3T}\biggr).                           
\end{eqnarray}
The low-frequency singularity in the scattering relaxation time
mentioned above is for electrons exactly at the Fermi surface, 
$\epsilon=\Delta$.
We note that for the electrons above the Fermi surface, $\epsilon>\Delta$
the singularity in the scattering relaxation time associated with the 
potential $V_{11}(Q)_{scatt}$ is weaker but it does not disappear.
More accurately such a divergence must be regularized
directly in the physically measurable quantity e.g. the tunneling
conductance. However it is not necessary because the contribution
to the scattering relaxation time from the phase collective mode, 
Eq. (43), is more important because it does not have a small
exponential factor such as that presented in Eqs. (42) and (45).

The appearance of the nonexponential scattering relaxation at low
temperature is a direct consequence of the gapless phase mode in
two dimensions. In three dimensions the phase mode have a gap
and the main contribution to the scattering relaxation comes
from the potential $V_{11}$.
\begin{eqnarray}
\label{46} 
{1\over \tau_{e-e}(T,\epsilon=\Delta)}_{scatt}=
{12\Delta T^{1/2}\over \pi(\pi D)^{3/2}\nu_3}  
\biggl({\Delta\over \omega_0}\biggr)^{1/4}
\exp\biggl(-{\Delta\over T}\biggr),                              
\end{eqnarray}
where $\nu_3=mp_F/\pi^2$ is the three-dimensional density of states.
Again after regularization of singularity in Eq. (46) we have
\begin{eqnarray}
\label{47}
{1\over \tau_{e-e}(T,\epsilon=\Delta)}_{scatt}=
{12T^{1/2}\over \pi(\pi D)^{3/2}\nu_3}  
\Delta\exp\biggl(-{4\Delta\over 5T}\biggr).         
\end{eqnarray}
The recombination relaxation time is obtained similar to Eq. (41),
\begin{eqnarray}
\label{48}
{1\over \tau_{e-e}(T,\epsilon=\Delta)}_{recom}=
{1\over 2^{1/4}(2\pi)^2} {\Delta^{1/2}T\over D^{3/2}\nu_3}
\biggl(1+\biggl({\lambda\nu_3\over 2}\biggr)^2\biggr)
\exp\biggl(-{2\Delta\over T}\biggr),                             
\end{eqnarray}

\section {Two-layer superconductor-normal metal system}

We consider a system of two disordered electron layers 
with different density of
states, $\nu_{1,2}$, elastic scattering times, $\tau_{1,2}$, 
mean free paths, $\ell_{1,2}$,
and diffusion coefficients $D_{1,2}$.
The layers are coupled by the Coulomb potentials, there is no
superconducting coupling between the layers. 

First we consider the screened Coulomb potentials in the normal state.
The nonscreened Coulomb potentials within the layer, $V_0$, and
between electrons in different planes, $U_0$, are
\begin{eqnarray}
\label{49}
V_0(q)=2\pi e^2/q\epsilon,\ \ \
U_0(q)={1\over \epsilon_1}{2\pi e^2\over q}\exp(-qb),         
\end{eqnarray}
where $\epsilon$ and $\epsilon_1$ are the dielectric constants of the
electron layer and the inter-layer media, $b$ is the distance between layers.
We assume that $\epsilon\approx\epsilon_1$ and we absorbed $\epsilon$ into
$e^2$

In all further calculations small momentum transfers are important, thus we
assume $qb<<1$,
\begin{eqnarray}
\label{50}
V_0-U_0=V_0(1-\exp(-qb))=2\pi e^2b,
\ \ \ V_0^2-U_0^2=V_04\pi e^2b.                           
\end{eqnarray}
In this chapter the lower indices of the potentials and the polarization
operators refer to the layer, e.g. $V_{11}$ means the Coulomb potential
between electrons in the layer 1, etc.
The screened potentials are satisfied the equations
\begin{eqnarray}
\label{51}
\pmatrix{V_{11}&U_{12}\cr U_{21}&V_{22}}=
  \pmatrix{V_0&U_0\cr U_0&V_0}                              
 +\pmatrix{V_0&U_0\cr U_0&V_0}\pmatrix{\Pi_1&0\cr 0&\Pi_2}
\pmatrix{V_{11}&U_{12}\cr U_{21}&V_{22}}.                         
\end{eqnarray}
We will use the definitions: $V_{11}=V_1$, $V_{22}=V_2$, and
$U_{12}=U_{21}=U$.
The solution of Eq. (3) is
\begin{eqnarray}
\label{52}
U={U_0\over P},\ \ 
V_1={V_0-(V_0^2-U_0^2)\Pi_2\over P}, \ \ \ 
V_2={V_0-(V_0^2-U_0^2)\Pi_1\over P},\cr
P=(1-V_0\Pi_1)(1-V_0\Pi_2)-U_0^2\Pi_1\Pi_2
\approx 1-V_0(\Pi_1+\Pi_2-4\pi e^2b\Pi_1\Pi_2).                   
\end{eqnarray}
The polarization operators in each layers for $q\ell_i<<1$ and 
$\omega\tau_i<<1$ are chosen in the form corresponding to a normal state,
because for recombination processes large frequencies $\omega>2\Delta$
are important, while collective excitations in a superconductor exist
only for $\omega<<\Delta$.
\begin{eqnarray}
\label{53}
\Pi^A_i(Q)=-\nu_i{D_iq^2\over i\omega+D_iq^2},        
\end{eqnarray}
The potentials are
\begin{eqnarray}
\label{54}
U^A(Q)={1\over \nu_1 (D_1+D_2\nu_2/\nu_1)q^2}
{(i\omega+D_1q^2)(i\omega+D_2q^2)
\over i\omega+\tilde Dq^2},                                        
\end{eqnarray}
\begin{eqnarray}
\label{55}
V^A_1(Q)=U^A(Q){i\omega+(1+2\kappa_2 d)D_2q^2\over i\omega+D_2q^2},
\end{eqnarray}                                                                 where
\begin{eqnarray}
\label{56}
\tilde D=\biggl(1+{\nu_2\over \nu_1}+2\kappa_2 d\biggr)
{D_1D_2\over D_1+D_2\nu_2/\nu_1},\ \ \  \kappa_2=2\pi e^2\nu_2.   
\end{eqnarray}  

We assume that layer 1 is in the superconducting state and layer 2 is in the
normal state. We will calculate the recombination relaxation time
in the superconducting layer due to inter-layer electron-electron
interaction $U$. From Eq. (40) we have
\begin{eqnarray}
\label{57} 
{1\over \tau_{e-e}(T,\epsilon=\Delta)}_{recom}=
{2\over \pi^3}\int_{2\Delta}^\infty 
d\omega\int_0^\infty dqq[N(\omega)+n(\omega-\Delta)]
{D_1q^2\over \omega^2-\Delta\omega +(D_1q^2)^2}\cr
{\rm Im}U^A(Q)\biggl({\Delta\over 2\omega}\biggl)^{1/2}.        
\end{eqnarray}  
For the imaginary part of the iner-layer interaction we use 
the approximation similar to Eq. (33), ${\rm Im}U^A(q,\omega)=
|U(q,0)|^2{\rm Im}\Pi_2^A(Q)$, where $U(q,0)$ corresponds to the static
limit of Eq. (53). The imaginary part of the polarization operator $\Pi_2$ 
corresponds to the electron scattering in the normal layer and thus
does not have a small exponential factor typical for a superconductor.
As a result
\begin{eqnarray}
\label{58}
{1\over \tau_{e-e}(T,\epsilon=\Delta)}_{recom}={T\over 4\pi^2\nu_1}
{D_1D_2\over (D_1+D_2\nu_2/\nu_1)^2}\biggl({1\over \tilde D}
+{1\over \tilde D+{\sqrt 2}D_1}\biggr)
\exp\biggl(-{\Delta\over T}\biggr).                              
\end{eqnarray}  
We see that relaxation rate in a superconductor-normal metal 
two-layer system is increased by an exponential factor in comparison with
the recombination rate in a single layer, see Eq. (40).

\section {Conclusions}

We derived the kinetic equation describing the electron relaxation in
two-dimensional impure superconductors. For electrons at the Fermi surface,
$\epsilon=\Delta$, the recombination and scattering relaxation time
were calculated for low temperatures, $T<<\Delta$. We took into account
all channels of the electron-electron interaction in the superconductor.

We found that the recombination relaxation rate comes from the
quasiparticle scattering (recombination processes) associated with 
the fluctuations of the electron density and the phase of the order
parameter, the propagators $V_{33}$ and $V_{22}$. The recombination 
relaxation rate has double exponential smallness at low temperatures
(see Eq. (41)),
associated with exponentially small number of available quasiparticles. 

The scattering relaxation rate has a power law temperature
dependence (see Eq. (43)) due to singularity in the propagators  $V_{22}$
and  $V_{33}$ associated with the gapless collective mode, the phase mode.
The contribution to the scattering relaxation rate from the fluctuations
of the amplitude of the order parameter, $V_{11}$ has an infrared divergence
similar to the phase relaxation time in the two-dimensional normal
metal \cite{AAK}, however after regularization the corresponding 
contribution to the scattering relaxation rate has a small exponential
factor and therefore is less important that the contribution from
the collective excitations.

We also shown that in the superconductor-normal metal two-layer
system the recombination
relaxation rate in the superconducting layer due to the inter-layer 
Coulomb interaction is strongly increased at low temperatures $T<<\Delta$
by an exponential factor $\exp(\Delta/T)>>1$ in comparison with 
a single superconducting layer. this fact may be important for
constructing superconducting radiation detectors \cite{SR} .

The author is grateful to I. L. Aleiner for valuable discussions
and A. V. Sergeev for his help at the early stage of the work.

\bigskip

\appendix 
\section{} 
\bigskip

In this Appendix we obtain equations for the polarization 
operators for some limiting cases and prove some identities for them.
Though some of the results were already presented in
Ref. 8, for Matsubara frequencies and in Ref. 15 for $T=0$ 
the analysis for continious frequencies has some advantages and helps
us to estimate the imaginary parts of the interaction propagators.

We start with the polarization operators for $q=0$.
$\Pi_{23}(0,\omega)$ may be taken directly from Eq. (21),
\begin{eqnarray}
\label{A1}
\Pi_{23}(0,\omega)=-{i\nu\omega\Delta\over 4}J(\omega),        
\end{eqnarray}
where
\begin{eqnarray}
\label{A2}
J(\omega)=\int{d\epsilon\over \xi_\epsilon\xi_{\epsilon+\omega}}
\biggl({S(\epsilon+\omega)-
S(\epsilon)\over \xi_{\epsilon+\omega}-\xi_\epsilon}
-{S(\epsilon+\omega)+
S(\epsilon)\over \xi_{\epsilon+\omega}+\xi_\epsilon}\biggr).      
\end{eqnarray}
Then we take $\Pi_{22}(\omega,0)$ from Eq. (19),
\begin{eqnarray}
\label{A3}
\Pi_{22}(0,\omega)={\nu\over 4}\int d\epsilon\biggl[
\biggl(1-{\epsilon(\epsilon+\omega)-\Delta^2\over 
\xi_\epsilon\xi_{\epsilon+\omega}}\biggr)          
{S(\epsilon+\omega)-S(\epsilon)\over \xi_{\epsilon+\omega}-\xi_\epsilon}+\cr
\biggl(1+{\epsilon(\epsilon+\omega)-\Delta^2\over 
\xi_\epsilon\xi_{\epsilon+\omega}}\biggr)          
{S(\epsilon+\omega)+S(\epsilon)\over \xi_{\epsilon+\omega}+\xi_\epsilon}
\biggr],                                                         
\end{eqnarray}
and transform it using the identities
\begin{eqnarray}
\label{A4}
1-{\epsilon(\epsilon+\omega)-\Delta^2\over 
\xi_\epsilon\xi_{\epsilon+\omega}}=
-{(\xi_{\epsilon+\omega}-\xi_\epsilon)^2-\omega^2
\over 2\xi_\epsilon\xi_{\epsilon+\omega}},                       
\end{eqnarray}
\begin{eqnarray}
\label{A5}
1+{\epsilon(\epsilon+\omega)-\Delta^2\over 
\xi_\epsilon\xi_{\epsilon+\omega}}=
-{(\xi_{\epsilon+\omega}+\xi_\epsilon)^2-\omega^2
\over 2\xi_\epsilon\xi_{\epsilon+\omega}},                       
\end{eqnarray}
to the form
\begin{eqnarray}
\label{A6}
\Pi_{22}(0,\omega)={\nu\over 4}\int d\epsilon
\biggl({S(\epsilon)\over \xi_\epsilon}+
{S(\epsilon+\omega)\over \xi_{\epsilon+\omega}}\biggr)+
{\nu\omega^2\over 8}J(\omega).                                  
\end{eqnarray}
Now recalling the BCS selfconsistancy equation
\begin{eqnarray}
\label{A7}
{2\over \lambda}+{\nu\over 2}\int d\epsilon
{S(\epsilon)\over \xi_\epsilon}=0,                              
\end{eqnarray}
we see that
\begin{eqnarray}
\label{A8}
{2\over \lambda}+\Pi_{22}(0,\omega)=
{\nu\omega^2\over 8}J(\omega).                                   
\end{eqnarray}
To transform $\Pi_{33}(0,\omega)$,
\begin{eqnarray}
\label{A9}
\Pi_{33}(0,\omega)+\nu=-{\nu\over 4}\int d\epsilon\biggl[
\biggl(1+{\epsilon(\epsilon+\omega)-\Delta^2\over 
\xi_\epsilon\xi_{\epsilon+\omega}}\biggr)          
{S(\epsilon+\omega)-S(\epsilon)\over \xi_{\epsilon+\omega}-\xi_\epsilon}+\cr
\biggl(1-{\epsilon(\epsilon+\omega)-\Delta^2\over 
\xi_\epsilon\xi_{\epsilon+\omega}}\biggr)          
{S(\epsilon+\omega)+S(\epsilon)\over \xi_{\epsilon+\omega}+\xi_\epsilon}
\biggr],                                                         
\end{eqnarray}
we use the identities
\begin{eqnarray}
\label{A10}
1+{\epsilon(\epsilon+\omega)-\Delta^2\over 
\xi_\epsilon\xi_{\epsilon+\omega}}={\xi_{\epsilon+\omega}-\xi_\epsilon\over
\omega}\biggl({\epsilon+\omega\over \xi_{\epsilon+\omega}}+
{\epsilon\over \xi_\epsilon}\biggr)-{2\Delta^2\over              
\xi_\epsilon\xi_{\epsilon+\omega}},                             
\end{eqnarray}
\begin{eqnarray}
\label{A11}
1-{\epsilon(\epsilon+\omega)-\Delta^2\over 
\xi_\epsilon\xi_{\epsilon+\omega}}={\xi_{\epsilon+\omega}+\xi_\epsilon\over
\omega}\biggl({\epsilon+\omega\over \xi_{\epsilon+\omega}}-
{\epsilon\over \xi_\epsilon}\biggr)+{2\Delta^2\over              
\xi_\epsilon\xi_{\epsilon+\omega}},                             
\end{eqnarray}
and get
\begin{eqnarray}
\label{A12}
\Pi_{33}(0,\omega)+\nu=-{\nu\over 2\omega}\int d\epsilon\biggl(
{\epsilon+\omega\over \xi_{\epsilon+\omega}}S(\epsilon+\omega)-
{\epsilon\over \xi_\epsilon}S(\epsilon)\biggr)+
{\nu\Delta^2\over 2}J(\omega).                                  
\end{eqnarray}
Now we recall another identity,
\begin{eqnarray}
\label{A13}
\int d\epsilon\biggl(
{\epsilon+\omega\over \xi_{\epsilon+\omega}}S(\epsilon+\omega)-
{\epsilon\over \xi_\epsilon}S(\epsilon)\biggr)=-2\omega,        
\end{eqnarray}
and get
\begin{eqnarray}
\label{A14}
\Pi_{33}(0,\omega)={\nu\Delta^2\over 2}J(\omega).             
\end{eqnarray}
We see from Eqs. (A5), (A8), and (A13) that the following identity holds 
\begin{eqnarray}
\label{A15}
\Pi_{33}(0,\omega)\biggl({2\over \lambda}+\Pi_{22}(0,\omega)\biggr)+
(\Pi_{23}(0,\omega))^2=0.                                        
\end{eqnarray}
At low temperatures,
$T<<\Delta$ and for small arguments $\omega<<\Delta$ and $Dq^2<<\Delta$,
the polarization operators are 
\begin{eqnarray}
\label{A16}
{2\over \lambda}+\Pi_{11}(Q)=
-\nu\biggl[1-{\omega^2\over 12\Delta^2}
+\pi{Dq^2\over 8\Delta^2}\biggr],
\end{eqnarray}
\begin{eqnarray}
\label{A17} 
{2\over \lambda}+\Pi_{22}(Q)=
-\nu\biggl[-{\omega^2\over 4\Delta^2}
+\pi{Dq^2\over 4\Delta^2}\biggr],
\end{eqnarray}
\begin{eqnarray}
\label{A18}
\Pi_{23}(Q)=-i\nu{\omega\over 2\Delta}, \ \ \
\Pi_{33}(Q)=-\nu\biggl[1+{\omega^2\over 6\Delta^2}\biggr].         
\end{eqnarray}
Thus for small arguments the following relation holds \cite{SRW}
\begin{eqnarray}
\label{A19}
\Pi_{33}(Q)\biggl({2\over \lambda}+\Pi_{22}(Q)\biggr)+
(\Pi_{23}(Q))^2\sim q^2.                                 
\end{eqnarray}

Using Eq. (A16) we have for the propagator $V_{11}$
\begin{eqnarray}
\label{A20}
V_{11}(Q)=-(2/\lambda+\Pi_{11}(Q))^{-1}
=\nu\biggl(1-{\omega^2\over 12\Delta^2}
+{\pi\over 8}{Dq^2\over \Delta^2}\biggr)^{-1}.                   
\end{eqnarray}
We see that $V_{11}$ is not singular which means that fluctuations
of the amplitude of the order parameter are massive, thus the 
imaginary part of the propagator $V_{11}$ originates from 
${\rm Im}\Pi_{11}(Q)$.

The screened Coulomb potential is presented in the form
\begin{eqnarray}
\label{A21}
V_{33}^A(Q)={V_0(q)\over 1-V_0(q)\tilde\Pi^A(Q)}, 
\end{eqnarray} 
where
\begin{eqnarray}
\label{A22}
\tilde\Pi^A(Q)=\Pi_{33}^A(q,\omega)+
{(\Pi_{23}^A(Q))^2\over 2/\lambda+\Pi_{22}^A(Q)}=
-\nu{\pi\Delta Dq^2\over \pi\Delta Dq^2-(\omega-i0)^2}.           
\end{eqnarray} 
The propagator $V_{22}$ may be written in a similar way
\begin{eqnarray}
\label{A23}
V_{22}^A(Q)=-{1\over {2\over \lambda}+\Pi_{22}^A(Q)}
{1-V_0(q)\Pi_{33}^A(Q)\over 1-V_0(q)\tilde\Pi^A(Q)}.
\end{eqnarray} 
The poles in the propagators $V_{33}$ and $V_{22}$
correspond to the collective
excitation, the phase mode, which in two dimensions is \cite{EP}
given by equation $\omega^2=\pi\Delta D\kappa q$. 

In quasi-one-dimensional superconductors the nonscreened
Coulomb potential is $V_0(q)=2e^2\ln(1/qa)$, $qa<<1$, where $a$ is
a cross-sectional size, and the density of states is $\nu_1=1/\pi v_F$,
thus the spectrum of the pase mode is \cite{MS}
\begin{eqnarray}
\label{A24}
\omega^2=\pi\Delta Dq^2\biggl({2e^2\over \pi v_F}\ln(1/qa)-1\biggr)
\end{eqnarray}

\section{}

It is interesting to see how the spectrum of the phase mode changes
in different two-layer systems. First we consider a system of two
identical impure superconducting planes coupled by the Coulomb interaction,
but no Josephson coupling between the planes, thus the order parameters
are independent in each planes.
For the screened Coulomb potential in each layer we have from Eq. (52)
\begin{eqnarray}
\label{B1}
V(Q)={V_0(q)-[V_0^2(q)-U_0^2(q)]\tilde\Pi(Q)\over 
[1-V_0(q)\tilde\Pi(Q)]^2-[U_0(q)\tilde\Pi(Q)]^2}.           
\end{eqnarray} 
To avoid confusion we dropped lower indices in the Coulomb potential $V$.
The spectrum of the phase modes is defined by equations
$1-(V_0+U_0)\tilde\Pi=0$ and $1-(V_0-U_0)\tilde\Pi=0$. the
solution of these equations for $qd<<1$
is $\omega_+=(2\pi\Delta D\kappa q)^{1/2}$ and
 $\omega_-=(2\pi\Delta D\kappa d)^{1/2}q$. These new phase modes
are similar to in-phase and out-of-phase plasmons in symmetric two-layer
clean normal metal system, see \cite{T} and \cite{DS}.

In a system of two coupled quasi-one-dimensional disordered superconductors
the spectrum of the phase modes in the long-wave limit $qd<<1$ is
\begin{eqnarray}
\label{B2}
\omega_+^2=\pi\Delta Dq^2\biggl({4e^2\over \pi v_F}\ln(1/qa)-1\biggr),\ \ 
\omega_-^2=\pi\Delta Dq^2\biggl({2e^2\over \pi v_F}\ln(d/a)-1\biggr).
\end{eqnarray}

Now we consider a two-layer superconductor-normal metal disordered system. 
The polarization operators in the superconducting and normal layers 
according to Eqs. (A22) and (53) are
\begin{eqnarray}
\label{B3}
\Pi_1^A(Q)=-\nu_s{\pi\Delta D_sq^2\over \pi\Delta D_sq^2-(\omega-i0)^2},
\end{eqnarray}                                                           
\begin{eqnarray}
\label{B4}
\Pi_2^A(Q)=-\nu_n{\pi D_nq^2\over i\omega+D_nq^2}.       
\end{eqnarray}
The spectrum of collective excitations is determined from the equation
\begin{eqnarray}
\label{B5}
1-V_0\biggl(\Pi_1+\Pi_2-4\pi e^2b\Pi_1\Pi_2\biggr)=0.         
\end{eqnarray}
For small momenta $q<<\kappa_s$, $\kappa_s=2\pi e^2\nu_s$ Eq. (B4) leads to 
\begin{eqnarray}
\label{B6}
\omega^2=\pi\Delta D_sq^2A +i\omega\pi\Delta{D_s\over D_n},\ \
A=1+{\nu_s\over \nu_n}+2\kappa_sb.
\end{eqnarray}
The solution of Eq. (B5) is a phase mode with small damping
\begin{eqnarray}
\label{B7}
\omega=(\pi\Delta D_sA)^{1/2}q+{i\over 2}\pi\Delta{D_s\over D_n},\cr
{\pi\over A}\biggl({D_s\over D_n}\biggr)^2<<{D_sq^2\over \Delta}<<1
\end{eqnarray} 
The last inequality is satisfied provided $A>>1$ and $D_s<<D_n$.
This result was independently obtained in \cite{NAA}. 
If the opposite inequality is valid, the solution of Eq. (B6) is
a diffusion mode,
\begin{eqnarray}
\label{B8}
i\omega+D_nq^2A=0,\ \  {D_sq^2\over \Delta}<<{\rm min}\biggl[
1,{\pi\over A}\biggl({D_s\over D_n}\biggr)^2\biggr].
\end{eqnarray}

\begin{figure}
\caption{$\Gamma(\hat\tau_i)$ is the impurity renormalized 
scalar vertex in the ladder approximation,  $V_{ij}$ is the
screened electron-electron interaction, $\Sigma$ is
the electron self-energy in an impure superconductor.} 
\end{figure}

\begin{table}
\caption{ Summary of the results for $\eta_i^{AA}$ and $\eta_i^{AR}$ defined
by equation Eq. (7). Quantities $A$, $A_+$, $B$, and $B_+$ are defined by
Eq. (9). }

\begin{tabular}{ccc}
$\eta_i$&$AA$&$AR$\\
\tableline
$\eta_0$&${\zeta_+\over 2}[(1-A_+)\hat\tau_0-B_+\hat\tau_1]$&
${\zeta\over 2}[(1+A_+)\hat\tau_0+B_+\hat\tau_1]  $\\
$\eta_1$&${\zeta_+\over 2}[(1+A_+)\hat\tau_1+B_+\hat\tau_0]  $&
${\zeta\over 2}[(1-A_+)\hat\tau_1-B_+\hat\tau_0] $\\
$\eta_2$&${\zeta_+\over 2}[(1+A)\hat\tau_2+iB\hat\tau_3] $&
${\zeta\over 2}[(1-A)\hat\tau_2-iB\hat\tau_3] $\\
$\eta_3$&${\zeta_+\over 2}[(1-A)\hat\tau_3+iB\hat\tau_2] $&
${\zeta\over 2}[(1+A)\hat\tau_3-iB\hat\tau_2] $\\
\end{tabular}
\end{table}

\begin{table}
\caption{ Summary of the results for impurity renormalized vertices 
$\Gamma_{22}^1(\hat\tau_i)$ and $\Gamma_{21}^2(\hat\tau_i)$.}

\begin{tabular}{ccccc}
$\Gamma(\hat\tau_i)$&$\hat\tau_0$&$\hat\tau_1$&$\hat\tau_2$&$\hat\tau_3$\cr
\tableline
$\Gamma^1_{22}(\hat\tau_1)$&$-\frac{\displaystyle\zeta}{\displaystyle 2}
\frac {B_+}{1-\zeta}$&
$1+\frac{\displaystyle \zeta}{\displaystyle 2}
\frac {\displaystyle 1-A_+} 
{\displaystyle 1-\zeta}$&$0$&$0$\\
$\Gamma^2_{21}(\hat\tau_1)$&$\frac{\displaystyle\zeta_+}{\displaystyle 2}
\frac {B_+}{1-\zeta_+}$&
$1+\frac{\displaystyle \zeta_+}{\displaystyle 2}
\frac {\displaystyle 1+A_+} 
{\displaystyle 1-\zeta_+}$&$0$&$0$\\
$\Gamma^1_{22}(\hat\tau_2)$&$0$&$0$&
$1+\frac{\displaystyle \zeta}{\displaystyle 2}
\frac {\displaystyle 1-A} {\displaystyle 1-\zeta}$&
$-i\frac{\displaystyle\zeta}{\displaystyle 2}\frac {B}{1-\zeta}$\\
$\Gamma^2_{21}(\hat\tau_2)$&$0$&$0$&
$1+\frac{\displaystyle \zeta_+}{\displaystyle 2}
\frac {\displaystyle 1+A} 
{\displaystyle 1-\zeta_+}$&
$i\frac{\displaystyle\zeta_+}{\displaystyle 2}\frac {B}{1-\zeta_+}$\\
$\Gamma^1_{22}(\hat\tau_3)$&$0$&$0$&
$-i\frac{\displaystyle\zeta}{\displaystyle 2}\frac {B}{1-\zeta}$&
$1+\frac{\displaystyle \zeta}{\displaystyle 2}
\frac {\displaystyle 1+A} 
{\displaystyle 1-\zeta}$\\
$\Gamma^2_{21}(\hat\tau_3)$&$0$&$0$&
$i\frac{\displaystyle\zeta_+}{\displaystyle 2}\frac {B}{1-\zeta_+}$&
$1+\frac{\displaystyle \zeta_+}{\displaystyle 2}
\frac {\displaystyle 1-A} 
{\displaystyle 1-\zeta_+}$\\
\end{tabular}
\end{table}

\begin{table}
\caption{Summary of the results for the polarization operators in the regions
of the parameters important for the electron relaxation at low temperatures
$T<<\Delta$.}

\begin{tabular}{ccc}
${\rm Im}\Pi^A_{ii}$&$\omega-2\Delta<<T<<\Delta$&$\omega<<T<<\Delta$\\
\tableline
$({\rm Im}\Pi^A_{22}={\rm Im}\Pi^A_{33})_{scatt}$&$\nu
\frac{\displaystyle (\pi\omega T)^{1/2}Dq^2\exp(-\Delta/T)}
{\displaystyle\omega^2+2\Delta\omega+(Dq^2)^2}$&
$2\nu \frac{\displaystyle\omega Dq^2\exp(-\Delta/T)}
{\displaystyle 2\Delta\omega+(Dq^2)^2}$\\
$({\rm Im}\Pi^A_{11})_{scatt}$
&$\nu\biggl(\frac{\displaystyle\pi T}{\omega}\biggr)^{1/2}
\frac{\displaystyle(2\Delta+\omega)Dq^2\exp(-\Delta/T)} 
{\displaystyle\omega^2+2\Delta\omega+(Dq^2)^2}$&
$\nu\biggl(\frac{\displaystyle\pi \omega}{T}\biggr)^{1/2}
\frac{\displaystyle 2\Delta Dq^2\exp(-\Delta/T)}
{\displaystyle 2\Delta\omega+(Dq^2)^2}$\\
$({\rm Im}\Pi^A_{22}={\rm Im}\Pi^A_{33})_{rec}$
&$\nu\frac{\displaystyle\pi \omega Dq^2}
{\displaystyle\omega^2-2\Delta\omega+(Dq^2)^2}$&$-$\\
$({\rm Im}\Pi^A_{11})_{rec}$&$\approx 0$&$-$\\
\end{tabular}
\end{table}

\end{document}